\documentclass[onecolumn, draftcls, 12pt, twoside]{IEEEtran}
\IEEEoverridecommandlockouts
\usepackage{graphicx}
\usepackage[cmex10]{amsmath}
\newcounter{subeqn} \renewcommand{\thesubeqn}{\theequation\alph{subeqn}}%
\makeatletter
\@addtoreset{subeqn}{equation}
\makeatother
\newcommand{\subeqn}{%
	\refstepcounter{subeqn}
	\tag{\thesubeqn}
}
\usepackage{url}  
\usepackage{balance}
\usepackage{cite}
\usepackage{amssymb}
\usepackage{hyperref}
\usepackage{amsmath}
\usepackage{algorithm}
\usepackage{algpseudocode}
\usepackage{subfigure}
\usepackage{array}

\newcommand{\bs}{\boldsymbol}

\usepackage{dblfloatfix}

\ifCLASSOPTIONcaptionsoff
\usepackage[nomarkers]{endfloat}
\renewcommand{\caption}[2][\relax]{\MYoriglatexcaption[#2]{#2}}
\fi

\newtheorem{proposition}{Proposition}

\begin{document}
\title{Dynamic Bandwidth Allocation and Edge Caching Optimization for Nonlinear Content Delivery through Flexible Multibeam Satellites}

\author{\IEEEauthorblockN{Thang X. {Vu}$^1$, Nicola Maturo$^1$, Symeon Chatzinotas$^1$, Joel Grotz$^2$, Tom Christophory$^2$, and Bj\"orn Ottersten}$^1$ \\
	\IEEEauthorblockA{$^1$Interdisciplinary Centre for Security, Reliability and Trust (SnT), University of Luxembourg, \\
		L-1855 Luxembourg. E-mail: \{thang.vu, nicola.maturo, symeon.chatzinotas, bjorn.ottersten\}@uni.lu \\
	$^2$SES Engineering, L-6815 Betzdorf, Luxembourg. Email: \{joel.grotz, tom.christophory\}@ses.com}
\thanks{This work is accepted to the IEEE ICC 2022.}
 
}


\maketitle
\pagenumbering{gobble}
\begin{abstract}
The next generation multibeam satellites open up a new way to design satellite communication channels with the full flexibility in bandwidth, transmit power and beam coverage management. In this paper, we exploit the flexible multibeam satellite capabilities and the geographical distribution of users to improve the performance of satellite-assisted edge caching systems. Our aim is to jointly optimize the bandwidth allocation in multibeam and caching decisions at the edge nodes to address two important problems: i)  cache feeding time minimization and ii) cache hits maximization. To tackle the non-convexity of the joint optimization problem, we transform the original problem into a difference-of-convex (DC) form, which is then solved by the proposed iterative algorithm whose convergence to at least a local optimum is theoretically guaranteed. Furthermore, the effectiveness of the proposed design is evaluated under the realistic beams coverage of the satellite SES-14 and Movielens data set. Numerical results show that our proposed joint design can reduce the caching feeding time by 50\% and increase the cache hit ratio (CHR) by 10\% to 20\% compared to existing solutions. Furthermore, we examine the impact of multispot beam and multicarrier wide-beam on the joint design and discuss potential research directions.

\end{abstract}

\begin{IEEEkeywords}
Edge caching, flexible multibeam satellite, resource allocation, successive convex approximation.
\end{IEEEkeywords}


\section{Introduction}

Efficient data distribution is one of the most important goals in 5G and beyond networks due to not only the proliferation of connected devices but also the increasing demand for data-hungry applications. The major contributor to these extraordinary evolution of the network traffics is the mobile data, which will grow $74$\% by $2023$ according to recent Cisco's forecasts \cite{Cisco}.
This traffic growth results from the increase of mobile handsets, e.g., tablets and smart phones, along with the booming of streaming services, such as YouTube and NetFlix. In addition, the popularity of new applications that require real-time interactions and the increasing video quality, i.e., Tactile internet, Virtual Reality and 4K videos etc., also contribute to the network traffics, which in turns puts more pressure on both the core and access segments. From the content provider perspective, this trend is economically interesting as they can achieve more benefits either through subscriptions or advertising. On the other hand, the telecom operators seem not getting sufficient benefit for upgrading their infrastructure to serve the newly data-hungry applications. Furthermore, they also have difficulty in accessing new  frequency bands to improve their wireless access and backhaul networks as the spectrum resource is scarce. 
These ignite a new trend for the telecom operators to create data distribution services through their own content delivery networks (CDNs), in which efficient edge caching will be implemented to serve contents to end-users. Since the CDN belongs to the telecom operator, efficient cross-layer optimization can be easily performed between the physical infrastructure and the network service to improve the system resource usage and the user quality of experience \cite{5G:cache}. 

Edge caching through satellites has recently been considered as an efficient way to send popular data to many CDNs over wide areas of interest. Unlike caching via terrestrial networks, in which the cached data must go through multiple hops and be sent to each CDN individually, caching via satellite backhauling can reach many caching nodes due to the large beam coverage of the satellite channels \cite{Satcomsurvey}. In addition, the very large bandwidth allows the satellite to transport very high-volume data in short time. The joint consideration of terrestrial and satellite systems for edge caching is studied in \cite{sat_wire_com_cache_2005, Kalan17}, which employs satellite backhauling as the complementary solution to the terrestrial network. The large coverage of the satellite beams are shown to be beneficial in improving the caching performance of the proxy servers. A similar setup is considered in \cite{Brinton13}, in which caching update rules  based on the estimated local and global content popularity are executed via multicasting-based satellite links. The authors of \cite{Satellite:sate:2000} propose a broad/multi-cast based satellite to improve the caching performance of the caches at the user end. Hybrid configuration of monobeam and multibeam satellites is studied in \cite{VuIJSCN,VuKa} in which the whole frequency bandwith is shared between the wide-beam and multi-beam transmission modes. It is noted that the above-mentioned works consider the traditional satellite payload in which the mono/multi-beam configuration are fixed, and hence unable to exploit the flexibility of the next generation multibeam satellites.\\

\emph{Contributions:}
In this work, we consider the flexible multibeam satellite for fetching popular contents to the edge nodes and focus on the off-line placement phase to optimize the average caching performance \cite{Vu2017a,sat_wire_com_cache_2005,Kalan17,Satellite:sate:2000}. Our contributions are as follows: 
\begin{itemize}
	\item We formulate the joint bandwidth allocation and caching decision problem to fully exploit the potentials of future flexible multi-beam satellites and the geographical distribution of users towards efficient edge caching algorithm. The proposed design jointly optimizes the frequency bandwidth allocated to each beam and the content indexes to be stored at each edge nodes. We consider two important metrics of the caching systems: i) cache feeding time minimization and ii) cache hits maximization.
	\item To tackle the non-convexity of the formulated joint optimization problem, we propose an iterative algorithm using the successive convex approximiation (SCA) method. The convergence to at least a local optimum of the proposed iterative algorithm is guaranteed.
	\item We evaluate the proposed joint design under realistic Movielens data set \cite{movielens} and the SES-14's coverage \cite{SES14} over the US East Coast. Numerical results show that the proposed joint design significantly improves both the caching time and cache hit ratio (CHR) performances compared with existing reference schemes. In addition, we discuss potentials issues in flexible satellite systems to further boost the caching performance.  
\end{itemize} 

The rest of the paper is organized as follows. Section~\ref{sec:architecture} provides technological enablers for caching over satellite. Section~\ref{sec:Algorithm} describes the system parameters and the proposed joint design.  Section~\ref{sec:Results} presents numerical results. Finally, Section~\ref{sec:Conclusion} provides discussions and concludes the paper.

\section{Flexible Multibeam Satellites: Technological Enabler for Satellite-assisted Edge Caching}\label{sec:architecture}
Traditional satellite architectures usually consist of two distinct configurations: a wide beam with very large coverage which is mainly used for broadcasting applications, or multiple small beams which are designed for broadband services. These application-oriented configurations are best suitable for dedicated services, however reduce the agility and flexibility in coping with dynamic and complex situations. The main reason of such low flexible reconfiguration is the high cost and delay for changing the payload operation in the conventional architecture. With the development of novel payload technology, digital transparent payload (DTP) \cite{Airbus} and active on-board antennas are practically feasible which allows concurrent services sharing the same radio resources (power, bandwidth) in an efficient and reconfigurable hybrid broadcast/broadband modes. The recent successful launch of the satellite SES-17 with the fully digital payload actualizes the deployment of fully flexible multibeam satellites.

\begin{figure}[!t]
	\begin{centering}
		\includegraphics[width=0.6\columnwidth]{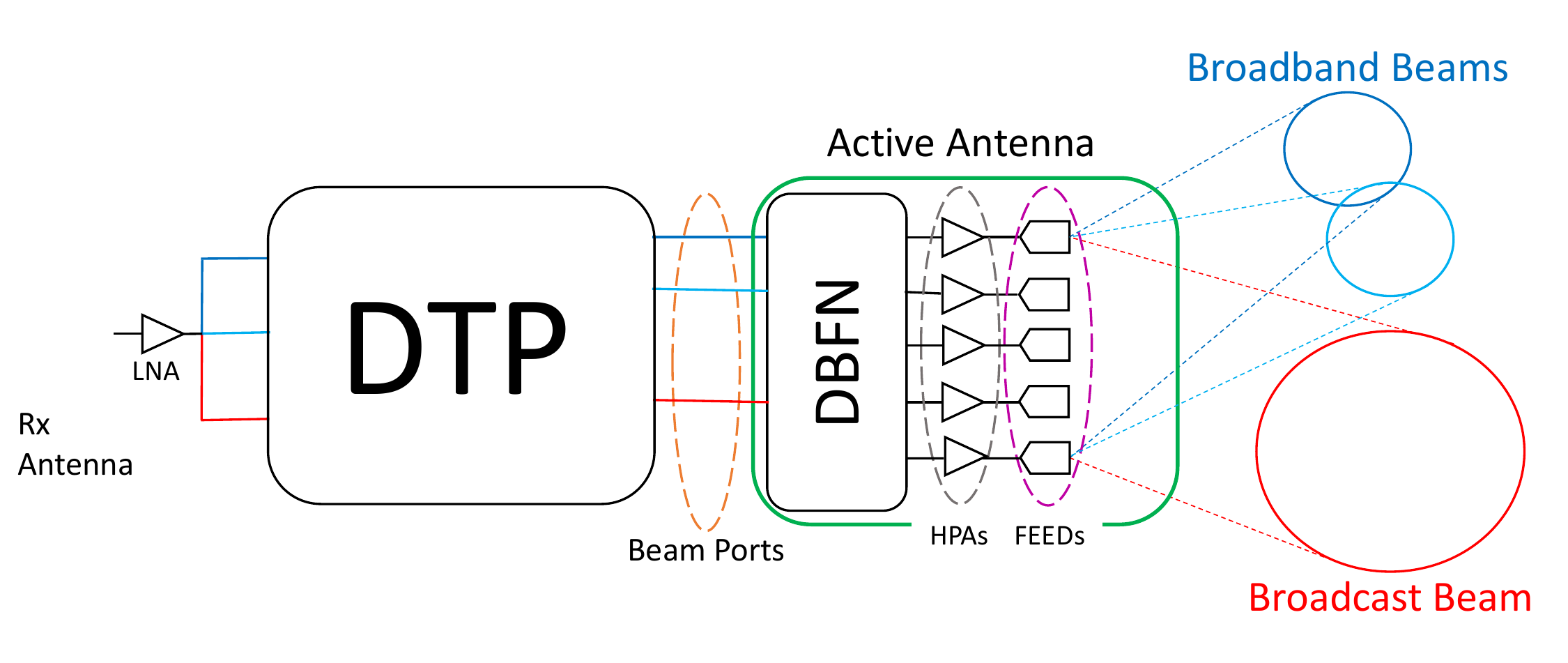}
		\caption{Illustration of the reference satellite coverage and architectures }
		\label{fig:flexiblepayload}
	\end{centering}
\end{figure}

Fig.~\ref{fig:flexiblepayload} depicts a pictorial example of flexible payloads sending three data streams: one broadcast stream and two broadband streams. By employing the active antenna technology, the three distinct data streams at the outputs of the DTP are simultaneously guided to the broadcast beam and two broadband beams from the same payload. More importantly, the flexible payload also enable arbitrary shape of the multibeams and concentrate power in the broadcasting beam to efficiently deal with non-uniform distribution of users. It is worth to note that this configuration does not require a direct connection between the beams and the feed. Instead, digital beamforming (BF) is employed by all the feeds to cooperatively generate the beams. This special architecture allows to optimize the energy consumption of the high power amplifiers (HPAs) serving the feeds. 
Furthermore, since the beams are generated by the digital beamforming network (DBFN), they are easy to modified and reconfigure. Such flexibility brings an extremely important feature to satellite operators to reduce controlling overhead in order to better tackle the variation of the traffic demands during the satellite lifetime. This is particularly crucial to edge caching systems, where the user preference (hence content popularity) largely varies cross geographical areas and over time.  

\section{Joint Caching and Bandwidth Allocation for Flexible Multibeam Satellites}\label{sec:Algorithm}
In this section, we develop a joint resource allocation and caching algorithm, which takes a further step to fully exploit the usage of the wide beam and multi-spot beam that share spectral and power resources through a DTP. In particular, the developed algorithm will jointly determine the caching decisions (files to go through which mode) as well as the required link rate of each mode, subjected to DTP resource constraints. Since our target is to optimize bandwidth to be used by both the multi-beam and the wide-beam modes, we calculate the link budget using the nominal bandwidth and power provided by SES. In addition, we assume that the power spectral density is constant, so that changing the bandwidth we are able to keep constant the signal-to-noise ratio (SNR). In this way the caching algorithm can optimize the bandwidth relying on a fixed SNR value (equivalently the spectral efficiency) for the location of interest.

The considered satellite-assisted edge caching system consists of $N$ CDN caches, which receive a selection of contents via both wide-beam and multi-spot beam coverages of the considered flexible satellite. Each cache serves the users within its coverage who are interested in contents from the common database of $F$ content files. We introduce variable $\bs{x} \in \{0,1\}^{F\times 1}$ presenting the indexes of files sent through the wide beam and variable $\bs{y}_n \in \{0,1\}^{F\times 1},n=1,…,N,$ presenting the indexes of files sent to the $n$-th cache through the multi-beam channels. In addition, toward intelligent caching we introduce variable $\bs{x}_n\in \{0,1\}^{F\times 1}, n = 1,\dots, N$ indicating the stored file indexes at the $n$-th cache from the wide beam coverage.
Let $\bs{l}_n \in \mathbb{N}^{F\times 1}, n=1,…,N$ denote the demands vector, in which the $f$-th component $\bs{l}_{n,f}$ is the average number of requests for the $f$-th content at the $n$-th CDN. 
In this off-line setting, the demand vectors are assumed to be obtained from the intelligent unit that predicts the content requests using, e.g., regression model or machine learning techniques\footnote{The joint design for on-line caching setting is left for future work and discussed in Section~\ref{sec:Conclusion}}. In simulation section, we obtain the demand vectors from the realistic Movielens data set. Therefore, the total number of hits at the CDN $n$ is calculated as 
$\bs{l}_n^T  (\bs{x}_n + \bs{y}_n ), \forall n$. Denote $\bs{q} = [q_1,…,q_f,…,q_F]^T$ as the vector of file sizes, in which $q_f$ is the size of the $f$-th content file.

\subsection{Minimization of the cache placement time}
In this subsection, we aim at minimizing the time duration, $\tau$, that the satellite spends for content fetching while satisfying the minimum cache hit requirements. The joint optimization of bandwidth allocation, cache feeding time, and caching decisions to every CDN is formulated as follows:
\begin{align}
	\underset{\bs{x}, \bs{x}_n, \bs{y}_n, \bs{w}, \tau}{\mathtt{Maximize}} ~~& \tau, \label{OP:2}\\
	\mathtt{s.t.} ~~
	& \bs{l}_n^T  (\bs{x}_n + \bs{y}_n ) \geq \eta_n, \forall n, \subeqn \label{eq:OP2 c1}\\
	& \bs{q}^T \bs{x} \leq w_0 \gamma_0 \tau, \subeqn \label{eq:OP2 c2} \\
	& \bs{q}^T \bs{y}_n \leq w_n \gamma_n \tau, \forall n, \subeqn \label{eq:OP2 c3} \\
	& \bs{q}^T(\bs{x}_n + \bs{y}_n) \leq M_n, \forall n, \subeqn \label{eq:OP2 c4}\\
	& \bs{x}_n + \bs{y}_n \leq \bs{1}, \forall n \subeqn \label{eq:OP2 c5} \\
	& \bs{x}_n \leq \bs{x}, \forall n, \subeqn \label{eq:OP2 c6}\\
	& {\sum}_{n=0}^N w_n \leq W_{tot}, \subeqn \label{eq:OP2 c7} \\
	& \bs{x}, \bs{x}_n, \bs{y}_n \in \{0,1\}^{F\times 1}, \forall n, \subeqn \label{eq:OP2 c8}
\end{align}
where $\bs{w} = \{w_n\}_{n=0}^N$ is the short-hand notation for the subcarriers, $M_n$ is the cache size of the $n$-th CDN, and $W_{tot}$ is the total available bandwidth. In the above optimization problem, constraint \eqref{eq:OP2 c1} is to guarantee the minimum cache hits requirement $\eta_n$ at every CDN; constraints \eqref{eq:OP2 c2} and \eqref{eq:OP2 c3} states that the total amount of sent data can not exceed the caching capacity of the wide-beam and multi-beam during the caching duration, respectively; \eqref{eq:OP2 c4} sets the maximum of cached data at each CDN not exceeding the cache size $M_n$; constraint \eqref{eq:OP2 c5} guarantees efficient caching and avoids one file being sent through both wide-beam and multi-beam; constraint \eqref{eq:OP2 c6} enables stored files from the wide-beam only if it is sent through the wide-beam; constraint \eqref{eq:OP2 c7} states that the total bandwidth of the wide-beam and multi-beam must not exceed the total bandwidth; finally the last constraint set binary values of corresponding decision variables.

Although having a linear objective function, solving problem \eqref{OP:2} is challenging because of constraints \eqref{eq:OP2 c2} and \eqref{eq:OP2 c3}.  In particular, the presence of products $w_0 \tau$ and $w_n \tau$ make constraints \eqref{eq:OP2 c2} and \eqref{eq:OP2 c3} non-convex. To tackle this difficulty, we will reformulate these two constraints into a preferable form which can be efficiently handled. By using the equivalent form $w_0 \tau = \frac{1}{2}((w_0+\tau)^2 - w^2_0 - \tau^2)$, constraints \eqref{eq:OP2 c2} is equivalent to following constraint:
\begin{align}
	\eqref{eq:OP2 c2} \Leftrightarrow 2\frac{\bs{q}^T \bs{x}}{\gamma_0} + w^2_0 + \tau^2 \leq (w_0 + \tau)^2. \label{eq:OP2 c2 app}
\end{align}
Because both sides of \eqref{eq:OP2 c2 app} are convex functions with respective variables, it suggests an successive approximation method to solve \eqref{OP:2} iteratively. In particular, let $\bar{w}_0, \bar{\tau}$ are feasible values of constraint \eqref{eq:OP2 c2 app} at iteration $i$-th. In the $(i+1)$-th iteration, the first-order approximation of the right-hand-side (RHS) of \eqref{eq:OP2 c2 app} at $\bar{w}_0, \bar{\tau}$ will be used, which is stated as follows:
\begin{align}
	2\frac{\bs{q}^T \bs{x}}{\gamma_0} + w^2_0 + \tau^2 \leq 2(w_0+\tau)(\bar{w}_0 + \bar{\tau}) - (\bar{w}_0 + \bar{\tau})^2, \label{eq:OP2 c2 app1}
\end{align}
which is convex because the left-hand-side (LFS) is a convex function and the RHS is linear. Similarly, we can approximate constraint \eqref{eq:OP2 c3} by the following convex constraint:
\begin{align}
	2\frac{\bs{q}^T\! \bs{y}_n}{\gamma_n}\! +\! w^2_n\! +\! \tau^2 \leq 2(w_n+\tau)(\bar{w}_n\! +\! \bar{\tau}) - (\bar{w}_n + \bar{\tau})^2. \label{eq:OP2 c3 app}
\end{align}
By using \eqref{eq:OP2 c2 app1} and \eqref{eq:OP2 c3 app} as the inner approximations of constraints \eqref{eq:OP2 c2} and \eqref{eq:OP2 c3}, respectively, and relaxing the binary constraint \eqref{eq:OP2 c8}, problem \eqref{OP:2} can be approximated by a convex optimization problem \eqref{OP:2 app}. Because of the nature of the approximation, the optimal solution of \eqref{OP:2 app} depends on the approximated points using in \eqref{eq:OP2 c2 app} and \eqref{eq:OP2 c3 app}. Therefore, to improve the optimal solution of the approximated problem \eqref{OP:2 app}, we propose an iterative Algorithm 1, which consists of a sequence of solving a convex optimization problem. The central principle of the proposed iterative algorithm is to have better initial values $\bar{\tau}, \bar{w}_n$ through iterations, which eventually reduces the gap between the approximated optimal solution and the original problem's solution. It is worthy mentioning that Algorithm 1 outputs continuous values of the caching decision variables $\bs{x}, \bs{y}_n, \bs{x}_n$. Therefore, a binary-recovery step will be performed at the end of Algorithm 1. Furthermore, to actively force the caching variables to be binary, a penalty function should be added to the objective function of \eqref{OP:2 app}. More details of penalty function can be found in \cite{Sajad21}.
\begin{algorithm}
	\caption{\textsc{Iterative Algorithm to solve (\ref{OP:2})}}
	\begin{algorithmic}[1]
		\State Initialize $\bar{\tau}, \bar{w}_n, \forall n$, $\epsilon$, $T_{\mathrm old}$ and $\mathtt{error}$.
		\State \textbf{while} $\mathtt{error} > \epsilon$ \textbf{do} 
		\State  \qquad Solve problem \eqref{OP:2 app} to obtain  $\tau^\star_k, \bs{x}^\star, \bs{x}^\star_n, \bs{y}^\star_n, w_n,  \forall n$
		\State \qquad Compute $\mathtt{error} = |\tau^\star - T_{\mathrm old}|$
		\State \qquad Update $T_{\mathrm old} \gets \tau^\star$;~ $\bar{\tau} \gets \tau^\star$; $\bar{w}_n \gets w^\star_n, \forall n$
	\end{algorithmic} 
\end{algorithm}
\begin{align}
	\underset{\bs{x}, \bs{x}_n, \bs{y}_n, \bs{w}, \tau}{\mathtt{Maximize}} ~~& \tau, \label{OP:2 app}\\
	\mathtt{s.t.} ~~
	& \eqref{eq:OP2 c1}, \eqref{eq:OP2 c4} - \eqref{eq:OP2 c7}, \eqref{eq:OP2 c2 app1}, \eqref{eq:OP2 c3 app}.\notag
\end{align}

\begin{proposition}\label{prop:E}
	The sequence of the objective values generated by Algorithm 1 in solving  problem \eqref{OP:2 app} is non-decreasing.
\end{proposition}
The proof of Proposition~\ref{prop:E} can be found by employing a technique similar to the one presented in \cite[Theorem 3.1]{DCProgram}, noting that both sides of constraints \eqref{eq:OP2 c2 app1} and \eqref{eq:OP2 c2 app} are continuously differential.
Proposition~\ref{prop:E} guarantees the convergence of Algorithm~1 to at least a (local) optimum. Although not guaranteeing the global optimality, it provides justification for the iterative algorithm.
\subsection{Cache Hits Maximization}
Although CHR is the popular metric of caching systems, it does not efficiently capture the non-uniform distribution of requests over different geographical areas. Therefore, we choose the cache hits as the performance metric of interest. We aim at jointly optimizing the caching decision and bandwidth allocation of the flexible multi-beam satellites to maximize the total number of cache hits at the CDNs.  The joint optimization is formulated as follows:
\begin{align}
	\underset{\bs{x}, \bs{x}_n, \bs{y}_n, \bs{w}}{\mathtt{Maximize}} ~~&{\sum}_{n=1}^N \bs{l}_n^T  (\bs{x}_n + \bs{y}_n ), \label{OP:1}\\
	\mathtt{s.t.} ~~& \eqref{eq:OP2 c2} - \eqref{eq:OP2 c8}, \notag
\end{align}
where the purpose of the constraints are presented in Section~\ref{sec:Algorithm}-A. 
Since both the objective function and constraints are linear, problem \eqref{OP:1} is a mixed binary linear problem (MBIP), which can be solved by either standard MBIP tools, e.g., Gurobi and Mosek, or relaxing the binary constraint with penalty function \cite{Sajad21}.
\section{Performance Evaluation on Realistic System Parameters}\label{sec:Results}
 In this section, we provide performance evaluation of the proposed joint caching and bandwidth allocation design, which exploits the flexible multi-beam satellite. The CHR performance shown in the figures are obtained by the cache hits obtained in Section~\ref{sec:Algorithm}-A divided by the total requests.

\subsection{Satellites Coverage}
\begin{figure}
	\centering
	\subfigure{\includegraphics[width=0.49\columnwidth]{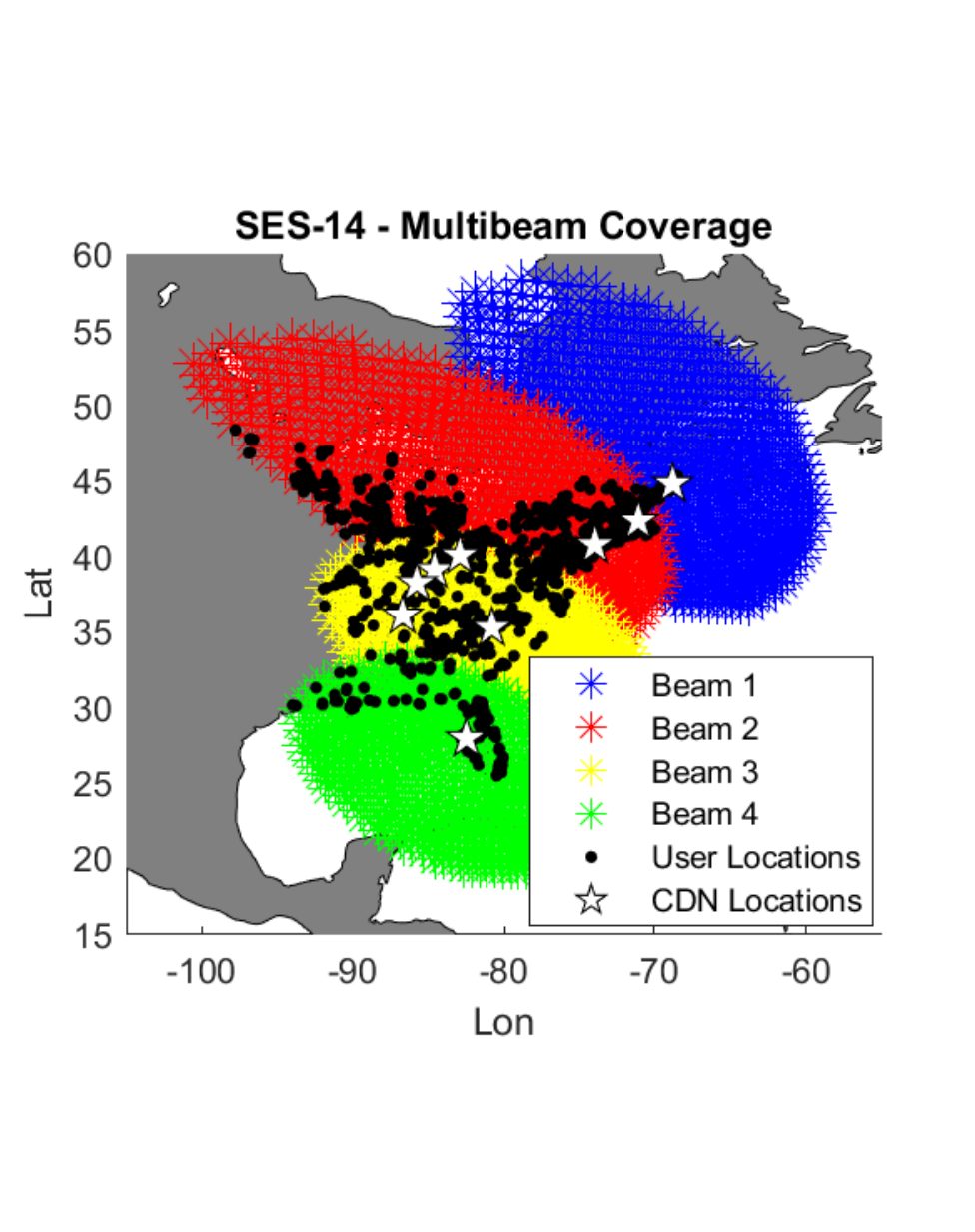}}
	\subfigure{\includegraphics[width=0.49\columnwidth]{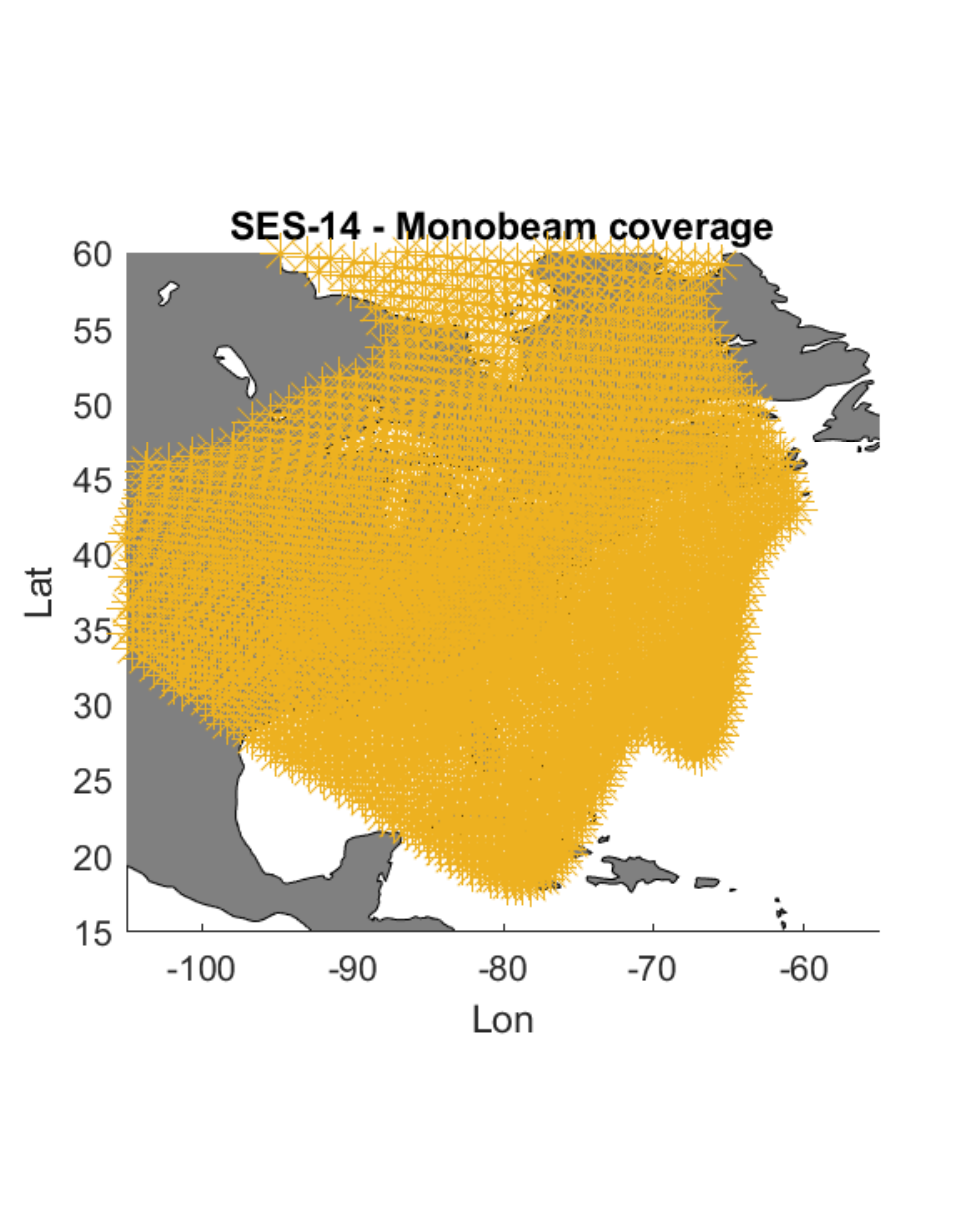}}
	\vspace{-1.1cm}
	\caption{SES-14 HTS KU multibeams (left) and monobeam (a) coverage on the US east coast. White starts are the CDNs' locations which are in major cities. Black dots are users' location, which are obtained from ZIP code information in the Movielens data set.} \label{fig:Coverage}
\end{figure}

In order to demonstrate the flexible frequency allocation between wide-beam and multi-beam modes, we employ the coverage of the satellite SES-14 \cite{SES14} on the US east coast as it provides both Ku-band HTS multi-beam footprint and single Ku-beam. 
Out of all the beams of the HTS system, we down select the 4 beams that cover more or less the same region of the US East monobeam, which are shown in Fig.~\ref{fig:Coverage}.
We used the power flux density information included in these beam patterns, and the per beam power information for calculating the signal to noise ratio (SNR) in any location within the coverage. The SNR at the CDNs' locations varies from $4.08$dB to $9.22$dB for the monobeam and from $2.48$dB to $9.33$dB for the multibeam.

\begin{figure}
	\centering
	\includegraphics[width=0.6\columnwidth]{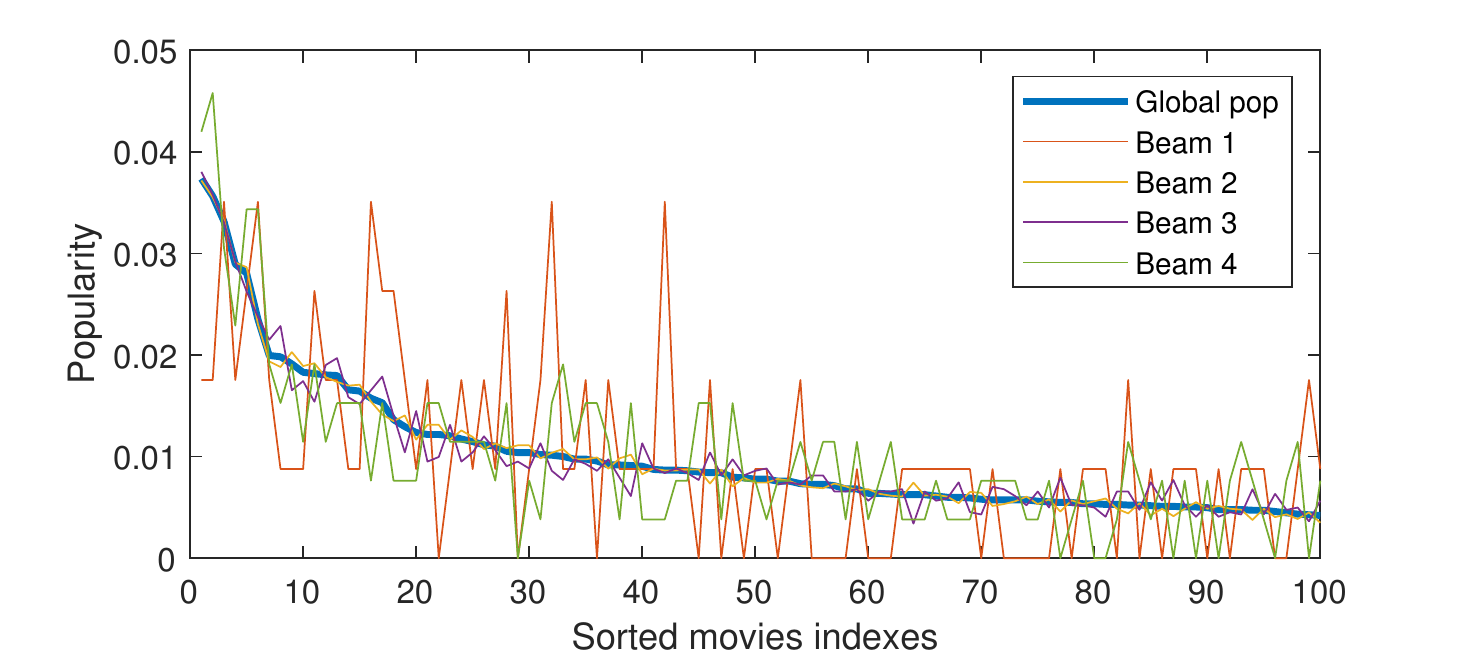}
	\vspace{-0.4cm}
	\caption{Content popularity of the considered files on different beams.} \label{fig:Pop}
		\vspace{-0.2cm}
\end{figure}

\subsection{Movielens-based Content Popularity}
To obtain realistic content popularity, we use the 1M data set made available by Grouplens \cite{movielens}, wherein 1 million movie ratings are provided. The data set provides user IDs, user locations (ZIP code), movie IDs, movie genres and the rating time, from which we can compute the distribution of requests (ratings) in arbitrary time duration. Then we use the Zip code information to exact geographical distributions of the requests, by mapping the Zip codes with corresponding  Latitude and Longitude components. Because the users in the 1M dataset locate all over the US territory, while the actual satellite's coverage are only the east coast, we only count the users within the satellite coverage when calculating local popularity  $\bs{l}_n$. Fig.~\ref{fig:Pop} shows the global and local (on each beam) popularities of 100 randomly selected movies. It is observed that the local popularity largely varies over different geographical areas.
\subsection{Performance Evaluations}
We conduct numerical results with 9 CDNs which are randomly selected from the satellite coverage. The global content popularity and local popularity on each beam are calculated based on the requests (user ratings in Movielens data set) from 100 movies which are randomly selected and have at least 100 ratings. The movie size is randomly varying between $0.5$ GBs and $1$ GBs. The cache size at each CDN varies from 5 GBs to 30 GBs. 

The proposed joint design is compared with three reference schemes where applicable: i) \emph{Reference 1} which is proposed in \cite{Brinton13} and use multi-beam to multicast the cached files; ii) \emph{Reference 2} which uses the whole frequency bandwidth to broadcast the globally popular files via the mono-beam; and iii) \emph{Reference 3} which is a hybrid mono/multi-beam based method \cite{Kalan17,VuKa} in which part of the bandwidth is used for broadcasting the globally popular files and the other part is used for multicasting the locally popular files to each CDN.  We remind that for a fair comparison, the total satellite bandwidth of all the schemes remain same.

\subsubsection{Cache feeding time performance}
\begin{figure}
	\centering
	\subfigure[Cache feeding time performance]{\includegraphics[width=0.6\columnwidth]{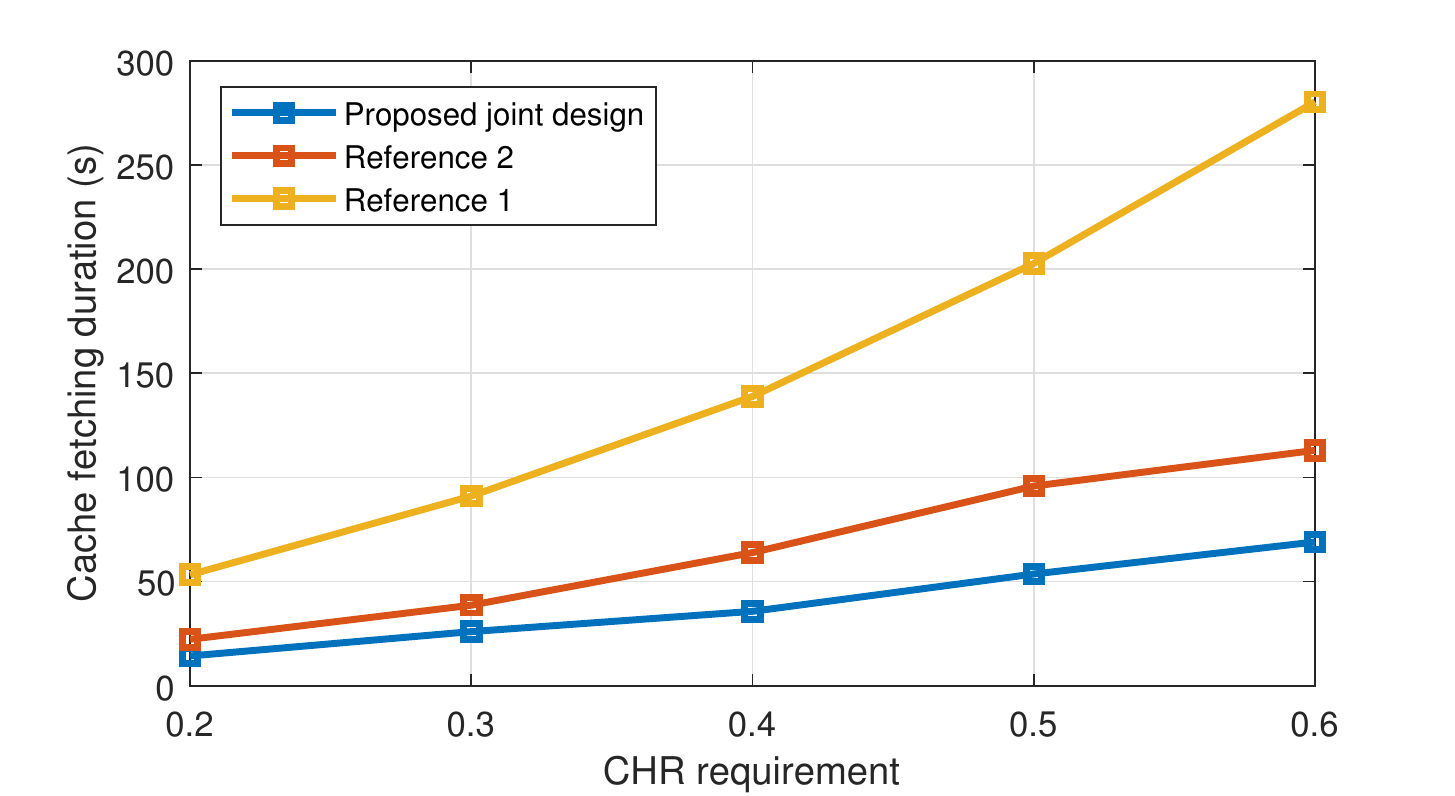}}
	\vspace{-0.4cm}
	\subfigure[Distribution of cached data]{\includegraphics[width=0.6\columnwidth]{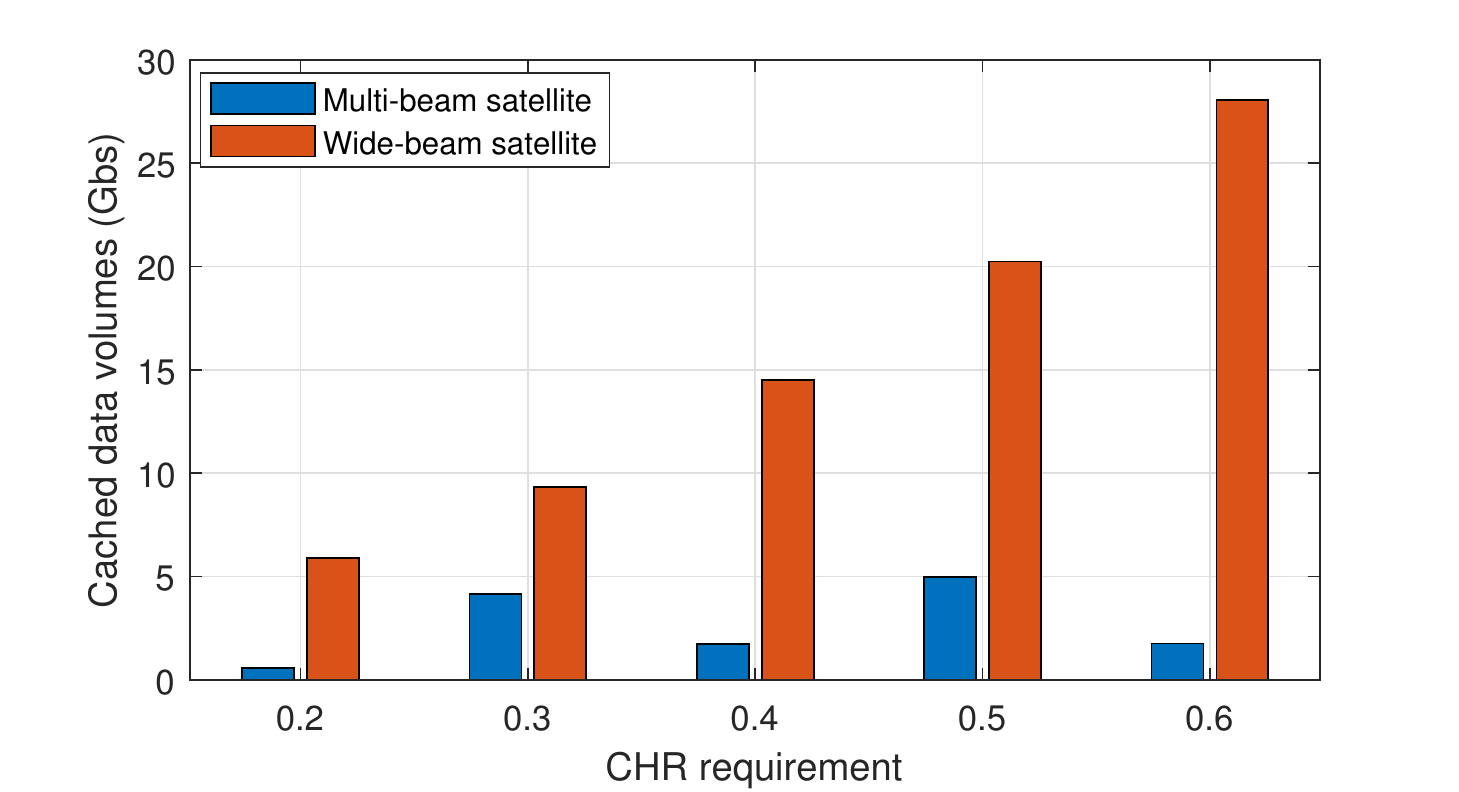}}	
	\caption{Caching feeding time comparison of the proposed design with Reference 1 and 2. The Reference 3 is not applicable in this case. The cache size is $30$ GBs.} \label{fig:Time}
\end{figure}

Fig.~\ref{fig:Time}a shows the time for caching as a function of the required CHR. We note that comparison with the Reference 3 is not available. In general, higher target CHR requires more cache fetching time as more files need to be sent to the caches. The advantage of the proposed joint design is clearly shown via the smallest caching time of the joint design compared to the references. In particular, at the target CHR of $0.5$, the proposed joint design reduces $70$\% and $50$\% the caching time compared to Reference 1 and Reference 2, respectively. An interesting observation is that Reference 2 (only uses the mono-beam) is more efficient than Reference 1 (only uses the multi-beam). This is because the caching time of Reference 1 is determined by the worst CDN while the resource allocated to each CDN is not optimized.
Fig.~\ref{fig:Time}b shows how the transferred data in the proposed joint design is distributed over different modes, i.e., via wide-beam and multi-beam. As the target CHR increases, more files are sent through the wide-beam because the most popular files are strongly correlated cross different geographical areas. 

\subsubsection{Cache hits performance}
\begin{figure}
	\centering
	\subfigure[Cache hit ratio comparison]{\includegraphics[width=0.6\columnwidth]{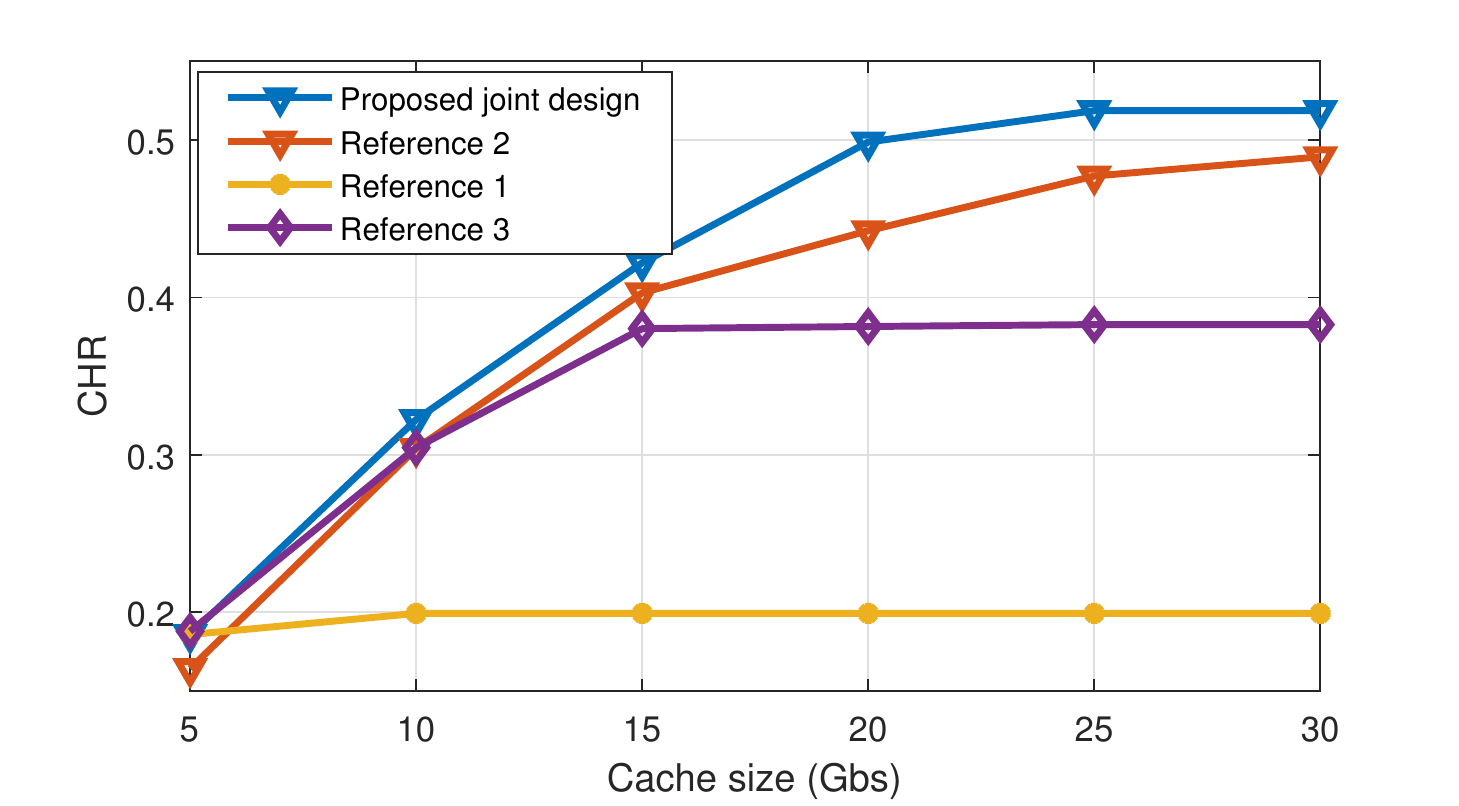}}
	\vspace{-0.4cm}
	\subfigure[Distribution of cached data]{\includegraphics[width=0.6\columnwidth]{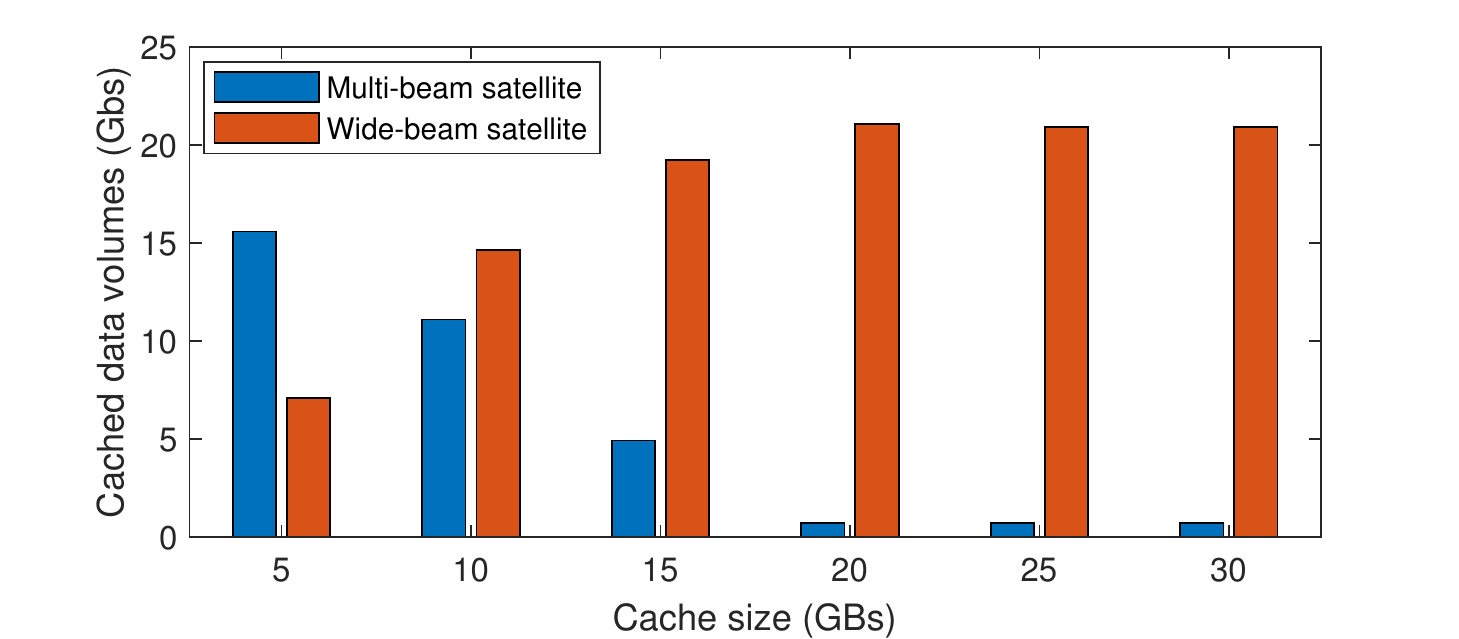}}
	\caption{CHR performance comparison. The caching time is equal to 100s.} \label{fig:CHR}
\end{figure}

In Fig.~\ref{fig:CHR}a, we compare the CHR performance of the proposed joint design with three reference schemes. It is observed that the proposed joint design achieves higher CHR compared to the references. This CHR gain comes from the efficient deployment of the multimodal capability of the satellite channel. In particular, two main factors determine the efficiency of the joint design: i) dynamic bandwidth allocation and ii) multimodal satellite channel. The first factor allows efficient bandwidth allocation to the beams with more demands. The second factor better serves some contents which are locally popular in some CDNs. 
Fig.~\ref{fig:CHR}b presents the cached data volumes allocation between the multi-beam and wide-beam modes of the proposed joint design as a function of the CDN’s cache size. When the cache size is small in relation to the database size, it is better to utilize both wide-beam and multi-beam modes. When the cache size is large, it can store most of contents sent by the satellites. In this case, using the wide-beam is sufficient.

\subsubsection{Multi-spot beam vs multicarrier widebeam}
\begin{figure}
	\centering
	\includegraphics[width=0.6\columnwidth]{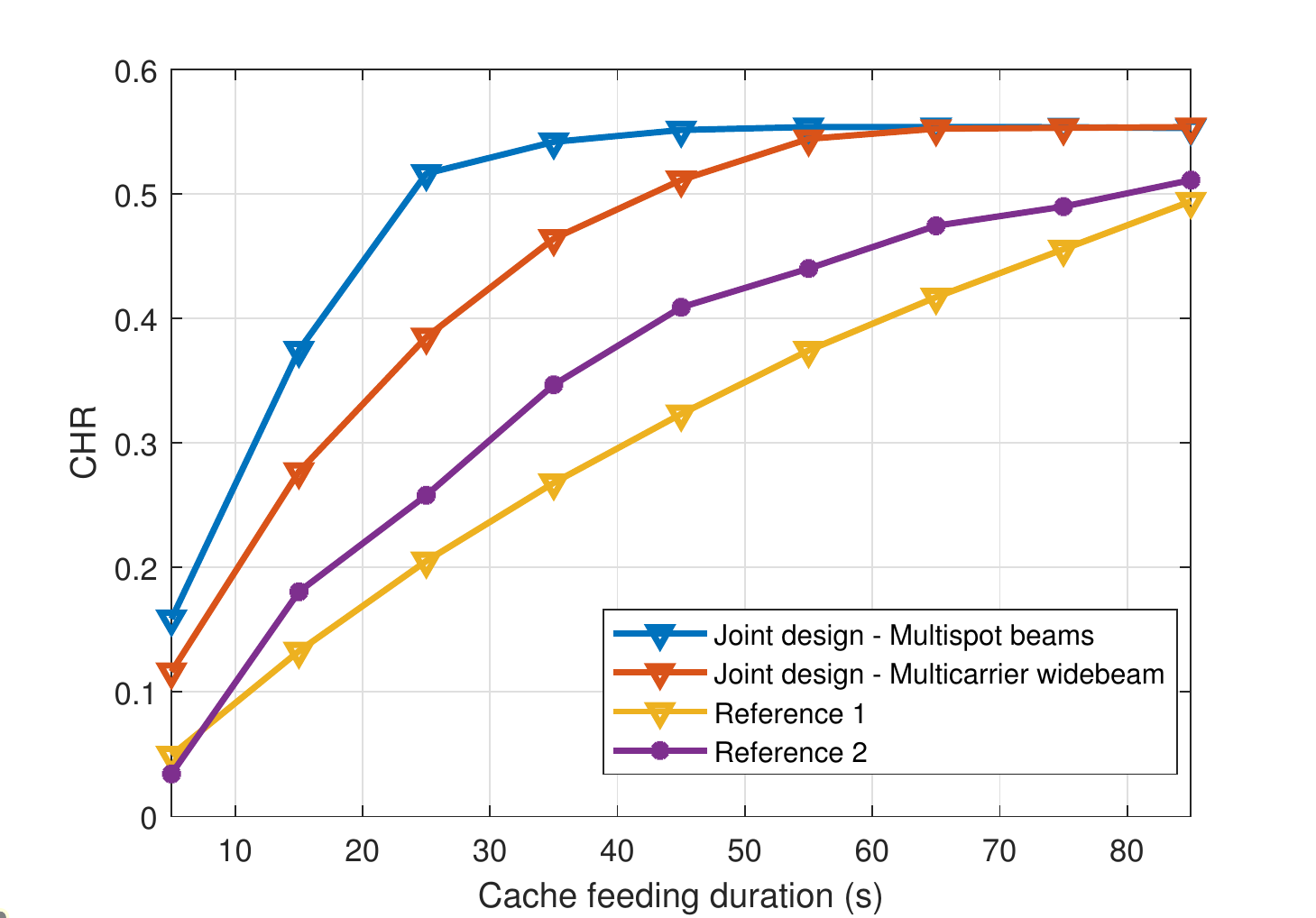}
	\vspace{-0.4cm}
	\caption{CHR performance of the joint design under multicarrier and multi-spot beam configurations. The cache size is 30 GBs.} \label{fig:multispot}
		\vspace{-0.2cm}
\end{figure}

In this subsection, we study different configurations of the satellite channel for implementing the multicasting mode in our joint design, namely the multi-spot beam based and the multicarrier wide-beam multicasting. In particular, due to different frequency coloring, frequency reuse can be obtained by the multi-spot beam mode. On the other hand, the multicarrier wide-beam is unable to reuse the frequency. We compare the CHR performance of the proposed joint design under both multi-spot beam and multicarrier wide-beam settings. There are four CDNs in this case, each CDN is served by one beam coverage. We assume that the frequency reuse factor is 2, which means that beam 1 and beam 3 share the same frequency sub-band, and beam 2 and beam 4 operate in the other sub-band. The cache size of the CDN is 30 GBs. The advantage of using the multi-spot beam is clearly shown in the Fig.~\ref{fig:multispot}, which outperforms the multicarrier when the cache feeding duration is limited. This gain results from the fact that due to frequency reuse, the multi-spot beam has more bandwidth for multicasting locally popular contents to the CDNs than the multicarrier wide-beam. When the caching time increases, the gain diminishes. This result is expected because a large cache feeding duration can deliver most of popular movies to the CDNs. In this case, additional locally popular movies brought by the multi-spot beam have negligible contribution to the overall CHR. 

We would highlight that the considered use case is the worst-case scenario and the performance gap is expected to increase with the multibeam system size. More specifically, the larger the available number of beams is, the more efficient it will become to use the multi-spot compared to the multicarrier framework. This can be intuitively explained by the fact that a wide-beam carrier renders the spectrum non-reusable for the entire coverage area, where a spot beam carrier can be reused multiple times depending on the number of supported spotbeams, e.g., the frequency reuse is equal to the number of beams divided by 4 for a 4C system. 
\section{Conclusion and Discussion}\label{sec:Conclusion}
In this paper, we have developed a joint caching and dynamic resource allocation that exploits the flexible multi-beam payload of the next generation multibeam satellites. In particular, we optimize the cached contents and the bandwidth allocation between the wide-beam and multi-beam modes in order to maximize the system cache hits. We use the realistic Movielens dataset to extract the content requests as well as the geographical distribution of requests. We demonstrated that the proposed joint design achieves about 10\% to 20\% higher cache hits compared to the reference under realistic request inputs. Furthermore, we demonstrated the benefit of using the multi-spot beam for multicasting to exploit the frequency reuse, compared with the multicarrier wide-beam setting. In addition, we provide a designing guidance for optimizing the system resource to maximize the CHR.

The current work assumes fixed power spectrum density, hence only the frequency bandwidth being optimized. Joint optimization of bandwidth, transmit power and beam coverage is expected to bring further performance gain at an expense of higher computation complexity. Another promising problem is to consider temporary content popularity. In this case, machine-learning based user preference prediction \cite{MyListofPapers:Bharat_TCOM_16,Lei18} will be jointly design the dynamic radio resource management of the flexible multibeam satellite. 
\balance

\section*{Acknowledgment}
We would like to thank Dr. Sumit Gautam for processing the Movielens data set. This work is partially supported by the Luxembourg National Fund (FNR) through the project FlexSAT under Grant C19/IS/13696663, and by the European Space Agency, ESA ESTEC, Noordwijk, The Netherlands, under the project HTS-DBS, contract No.  4000120692/17/NL/EM. Opinions, interpretations, recommendations and conclusions expressed herein are those of the authors and are not necessarily endorsed by the European Space Agency.


\end{document}